\documentclass[lettersize,journal]{IEEEtran}
\usepackage{amsmath,amsfonts}
\usepackage{algorithmic}
\usepackage{algorithm}
\usepackage{array}
\usepackage{xcolor}
\usepackage[caption=false,font=normalsize,labelfont=sf,textfont=sf]{subfig}
\usepackage{subcaption}
\usepackage{textcomp}
\usepackage{stfloats}
\usepackage{url}
\usepackage{verbatim}
\usepackage{graphicx}
\usepackage{url}
\usepackage{cite}
\begin{document}

\author{
\IEEEauthorblockN{Chris Albert\IEEEauthorrefmark{1,2}, Ritoban Basu Thakur\IEEEauthorrefmark{2,1}, Farzad Faramarzi\IEEEauthorrefmark{2}, Byeong Ho Eom\IEEEauthorrefmark{2}, Sumit Dahal\IEEEauthorrefmark{3}, Andrew Bear\IEEEauthorrefmark{4}, Reinier Janssen\IEEEauthorrefmark{2,1}, Henry LeDuc\IEEEauthorrefmark{2}, Thomas Stevenson\IEEEauthorrefmark{3}, and Peter Day\IEEEauthorrefmark{2}
}
\\
\IEEEauthorblockA{\IEEEauthorrefmark{1}California Institute of Technology, 1200 E California Blvd, Pasadena, CA 91125, United States of America}\\
\IEEEauthorblockA{\IEEEauthorrefmark{2}Jet Propulsion Laboratory, 4800 Oak Grove Dr, Pasadena, CA 91011, United States of America}\\
\IEEEauthorblockA{\IEEEauthorrefmark{3}Goddard Space Flight Center, 8800 Greenbelt Rd, Greenbelt, MD 20771, United States of America}\\
\IEEEauthorblockA{\IEEEauthorrefmark{4}University of Washington St. Louis, One Brookings Drive, St. Louis, MO 63130, United States of America}

}



\title{In Situ Interferometric Spatial Mapping Of A Microwave Kinetic Inductance Detector Array}
\maketitle

\begin{abstract}
We present a method of spatially mapping microwave kinetic inductance detector (MKID) arrays, in a dark setup. MKIDs are superconducting natively multiplexed resonators which enable kilopixel arrays, such as for the proposed Probe far-Infrared Mission for Astrophysics (PRIMA). In such telescope applications one must map the spatial location of each MKID with their individual resonance frequencies. Traditional LED arrays or beam-mapping methods become increasingly difficult as pixel spacing decreases, e.g., 900~$\mu$m separated MKIDs in the spectrometer module of PRIMA. Our new mapping technique uses a cryogenic interferometer in reflection mode. As on-resonance signals reflect from an MKID, they accrue a phase proportional to the path-length, exactly corresponding to their physical distance on the feedline. Specifically, we use a superconducting transmission line that has nonlinear kinetic inductance. The slow-wave structure of this nonlinear device is designed to have a signal speed of 0.64\% the speed of light, enabling a compact system.  Current biasing this line allows for varying the wave speed and ensuring that the phase measured is periodic within a nulling interferometric mode. Using this setup, we measure a length ordering that reflects the bimodal MKID distribution of a 44 pixel array of MKIDs designed for PRIMA which contains the same spacing as the final kilopixel array design.
\end{abstract}


\section{Introduction}

\IEEEPARstart{I}{n} focal plane arrays, MKIDs are strung together on one common feedline to enable frequency multiplexed readout\cite{JZ12, JB17, LF24}. Given specific design considerations, there may be no strict correspondence between an MKID's resonance frequency (\( f_{res}\)) and its physical location, say, distance from one of the feedline ports ($D$). We define the set of resonance frequencies as $\mathbf{F} = \{  f_{res,i}, \; i = 1 \ldots N_r \}$ with explicit ordering $f_{res,i} < f_{res,i+1}$ ($N_r$ is the total number of measured resonances). Similarly, $\mathbf{D}=\{D_i, \; i = 1 \ldots N_r\}$ where $D_i$ is the distance from port-1 where a resonator, whose resonance frequency is $f_{res,i}$, is located. We aim to extract the mapping between frequency and distance sets, $\mathbf{F} \mapsto \mathbf{D}$ for any array, and we present preliminary efforts to map an array with frequencies from 584 MHz to 1.1 GHz with 20 cm of feedline between the first and last resonator.

\section{Method}
Fig.~\ref{fig:setup} (a) shows our experimental setup. The ``beam splitter'' of our interferometer is a commercial 90$^{\circ}$ or quadrature hybrid coupler (QHC). An adjustable and pre-calibrated phase shifter (PS) is critical to perform nulling interferometry. This is shown as the ``$\phi(I)$'' element, a current biased transmission line. This line has a 0 $\Omega$ reflector as termination and forms one arm of the interferometer. In the other arm we have our MKID array labeled as device under test (DUT), terminated with 50 $\Omega$ for cryogenic power dissipation. The other two ports of the QHC are connected to a commercial Vector Network Analyzer (VNA). In essence, signal on resonance for some MKID $k$ will be reflected from the DUT and will interfere with a signal of the same frequency coming from the PS, but with a user controlled phase. We ramp the phase and find values where there are constructive and destructive (nulls) fringes. These specific values relate the phase in the signal coming from the DUT, in which the physical distance is encoded.

\subsection{Nonlinear Kinetic Inductance \& Interferometry}

We utilize a current controlled phase shifter (PS), a thin-film NbTiN transmission line where the kinetic inductance is a nonlinear function of the supercurrent due to a modified density of states and reduced energy gap \cite{AA03}. The phase shifter's design is detailed in \cite{FF25}, including the microstrip slow-wave structure that results in a propagation velocity of $0.0064c$. A similar design for use at higher frequencies is also presented in \cite{FF24}. The (DC) current controlled delay is \cite{BHE12}

\begin{equation}
\label{eq:delay}
    \tau(I) = \sqrt{\mathcal{L}(I) \mathcal{C}} = \tau_0 \sqrt{1 + (I/I_2)^2 + (I/I_4)^4 \ldots }
\end{equation}

The characteristic currents $I_2$ and $I_4$ can be found by fitting to the following relation between the current dependent frequency shift of the phase shifter's band gap and the nonlinear kinetic inductance \cite{SS21}. 

\begin{equation}
\label{eq:nonlinear_L}
    \left(\frac{f_{\text{gap}}(0)}{f_{\text{gap}}(I)}\right)^2 = \frac{\mathcal{L}(I)}{\mathcal{L}(0)} = 1 + \left(\frac{I}{I_2}\right)^2 + \left(\frac{I}{I_4}\right)^4
\end{equation}

\begin{figure}[h!]
\centering
{\includegraphics[width=0.9\linewidth]{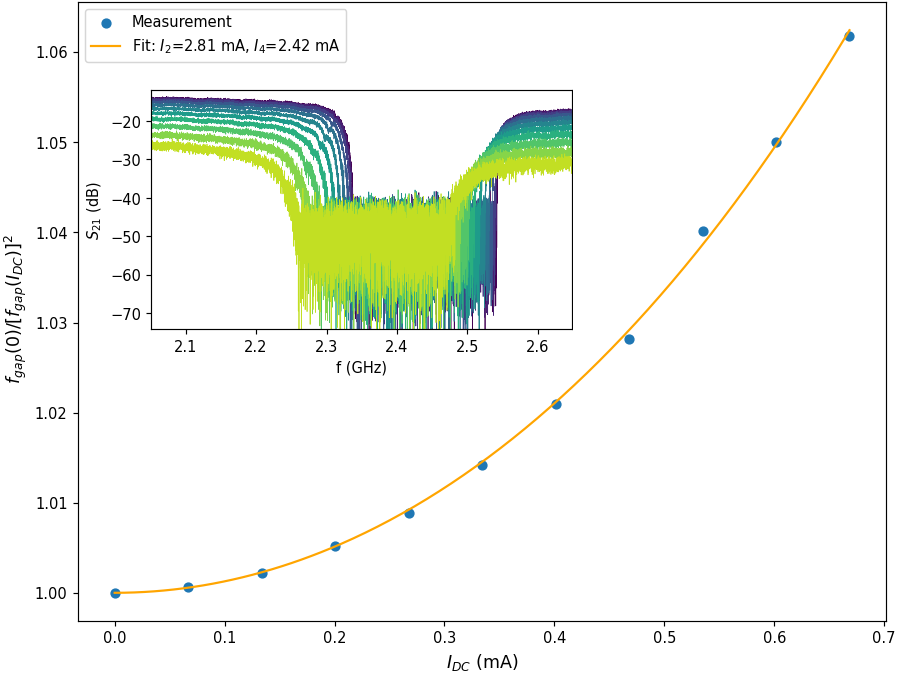} }
\caption{ Band gap shift of the phase shifter as a function of applied DC current and the fit to Eqn. \ref{eq:nonlinear_L}. \textbf{Insert:} The band gap in the phase shifter's transmission for each applied current. }
\label{fig:bandgap}
\end{figure}

Fig. \ref{fig:bandgap} shows Eqn. \ref{eq:nonlinear_L} fit to the phase shifter, where we find $I_2$ = 2.81 mA and $I_4$ = 2.42 mA. This agrees with the values found for the TiN device in \cite{FF25} when corrected by the ratio of critical temperatures between TiN and NbTiN. 

A phase shift can be dialed in as $\phi(I) = 2 \pi f \tau (I)$ for any propagating wave of frequency $f$.  An interferometric set up with the PS in one arm and DUT in the other is  shown in Fig.~\ref{fig:setup}. The reflected signal from both are combined with the quadrature hybrid coupler. This combined signal forms an interferogram w.r.t to delay, more in Eqn.~\ref{eq:P_out}. 
\subsection{Setup \& Calculation}

\begin{figure}[h!]
\centering
\subfloat[]{\includegraphics[width=0.9\linewidth]{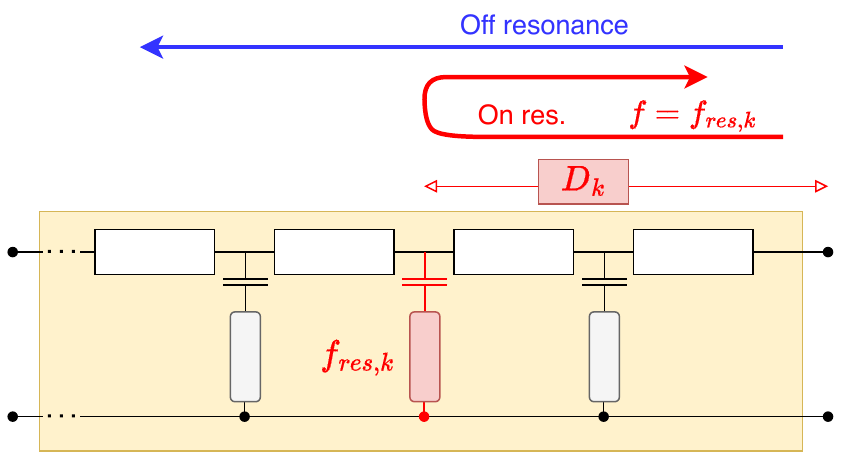} }
\label{fig:subim1}
$\quad$
\subfloat[]{\includegraphics[width=0.9\linewidth]{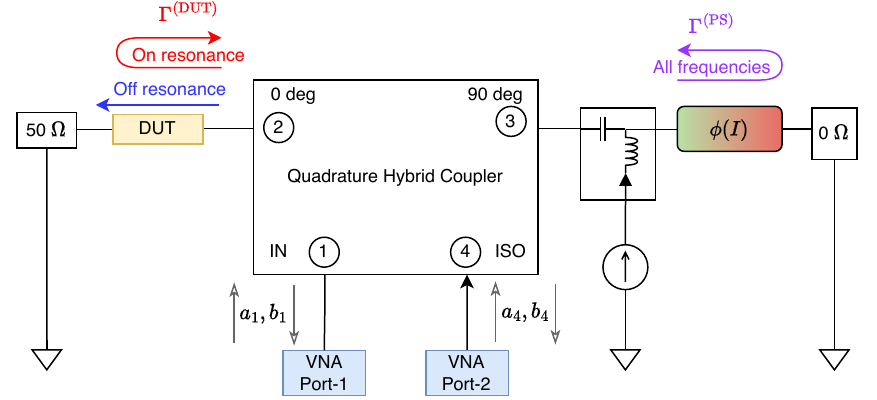} }
\label{fig:subim2}
\caption{Experimental setup \textbf{(a)} with DUT zoom in \textbf{(b)}. Delay of reflected signal from DUT is $T_k \equiv D_k/v$. Distance to $k^{th}$ KID from DUT input is $D_k$, wave-speed is $v$. Delay from PS is Eqn.~\ref{eq:delay}.}
\label{fig:setup}
\end{figure}


For input voltage $a_1$, the output is $b_4$, Eqn.~\ref{eq: b4_ideal}\cite{Pozar12} (Ref. Fig.~\ref{fig:setup} (right)).
\begin{equation}
\label{eq: b4_ideal}
    b_4 (f,I) = \frac{ia_1}{2}\left( \Gamma^\text{(DUT)}(f) + \Gamma^\text{(PS)}(f,I) \right),
\end{equation}
where the subscripts denote the port number on the QHC. 

Consider resonator $k$, with frequency $f_{res,k}$ at a distance  $D_k$ from the DUT input port. On resonance signals ($f = f_{res,k}$) are reflected, and interfere with the PS's reflection, combined and sensed at Port 4 of the coupler. For brevity we define $T_k \equiv D_k/v$ as wave-speed is not strictly required for finding the distance ordering. Similarly we define $\Omega_k = 4\pi f_{res,k}$ (known term). The extra factor of 2 is due to double travel on reflection.  

In the ideal limit, $\Gamma^\text{(DUT)}(f=f_{res,k}) \approx \text{exp}[i\pi + i\Omega_kT_k]$ (one bounce-back). The PS always reflects back the signal and $\Gamma^\text{(PS)} = \text{exp}[i\pi + i \Omega_k \tau(I)]$. The normalized power measured by the VNA gives us an interferogram, Eqn.~\ref{eq:P_out}. Note that we are using current implicitly. At Port 4 of the QHC, the measured power when on-resonance and in the ideal limit is\cite{Pozar12}

\begin{align}
\label{eq:P_out}
    P(\tau) & =   \frac{|b_4|^2}{|a_1|^2} = \frac{1}{4}{| \Gamma^\text{(PS)}+\Gamma^\text{(DUT)}|}^2 \\
    & \propto \frac{1}{2} \left(1 + \cos \left[\Omega_k(T_k - \tau)  \right]\right)
\end{align}

Interferograms can be fit with $\text{Fit}(\tau) = p_o + p_a \cos(p_b\tau + p_c)$. Parameters $p_o$ and $p_a$ set the contrast and can be ignored. For resonator $k$, we have $p_{b,k} = - \Omega_k$ (known) and fitting  acts as a consistency check. Crucially, $p_{c,k} = p_bT_k + \phi_0$ contains $T_k$ (distance term), and can have an additional offset $\phi_0$. As long as variations in $\phi_0$ from resonator to resonator are modest\footnote{For the smallest distance of 0.9mm, the variance in $\phi_0$ should be $< \text{0.9mm}\times 4 \pi \times 1 \text{GHz}/(0.4c) \approx $0.09 radians.}, we have a linear mapping between $p_{c,k}$ and $T_k$ consistent from resonator to resonator. Thus a robust ordering of resonators in physical space is attainable as shown below.

\begin{equation} \label{fitstuff1}
\begin{split}
    \text{Fit}_k(\tau) & = p_{o,k} + p_{a,k} \cos \left(p_{b,k}\tau + p_{c,k} \right) \\
 p_{b,k} &= - \Omega_k \propto -f_{res,k}; \quad \text{consistency-check} \\
 p_{c,k} &=  \phi_0 + p_{b,k} {T_k}
\end{split}
\end{equation}

\section{Some considerations}
\subsection{Can we resolve practical distances with this method?}

The minimum KID-KID distance is $D_k \approx$  0.9 mm, DUT wave speed is $v \approx 0.4c$, estimated from the geometry of the coplanar waveguide feedline\cite{simons01}. So, $T_k \approx 7.5$ ps. For the PS, we estimate delay noise via current noise (the minimum delay possible). The AFG31102 voltage source as was used, which has 0.1 mV RMS noise\footnote{\url{https://www.tek.com/en/datasheet/arbitrary-function-generators}}. We biased via a 10 k$\Omega$ resistor and $\sigma_I \approx $ 10 nA. Delay uncertainty is $\sigma_\tau \approx \tau_0I\sigma_I/I_2^2 \approx (238 ns)(1 mA)(10nA)/(3mA)^2 = $ 0.26 ps. As long as $\sigma_I<$200 nA, this interferometric process should work, and our measurements are not limited by such uncertainties.

At most, $T_k\approx 20 \text{cm}/(0.4c) = $1.7 ns. The PS we are using has $\tau \in (238, 246)$ ns for $I \in (0,0.7)$ mA, and therefore has enough range. In summary, this method is practically realizable.

\begin{figure}[h!]
    \centering
\includegraphics[width=1\linewidth]{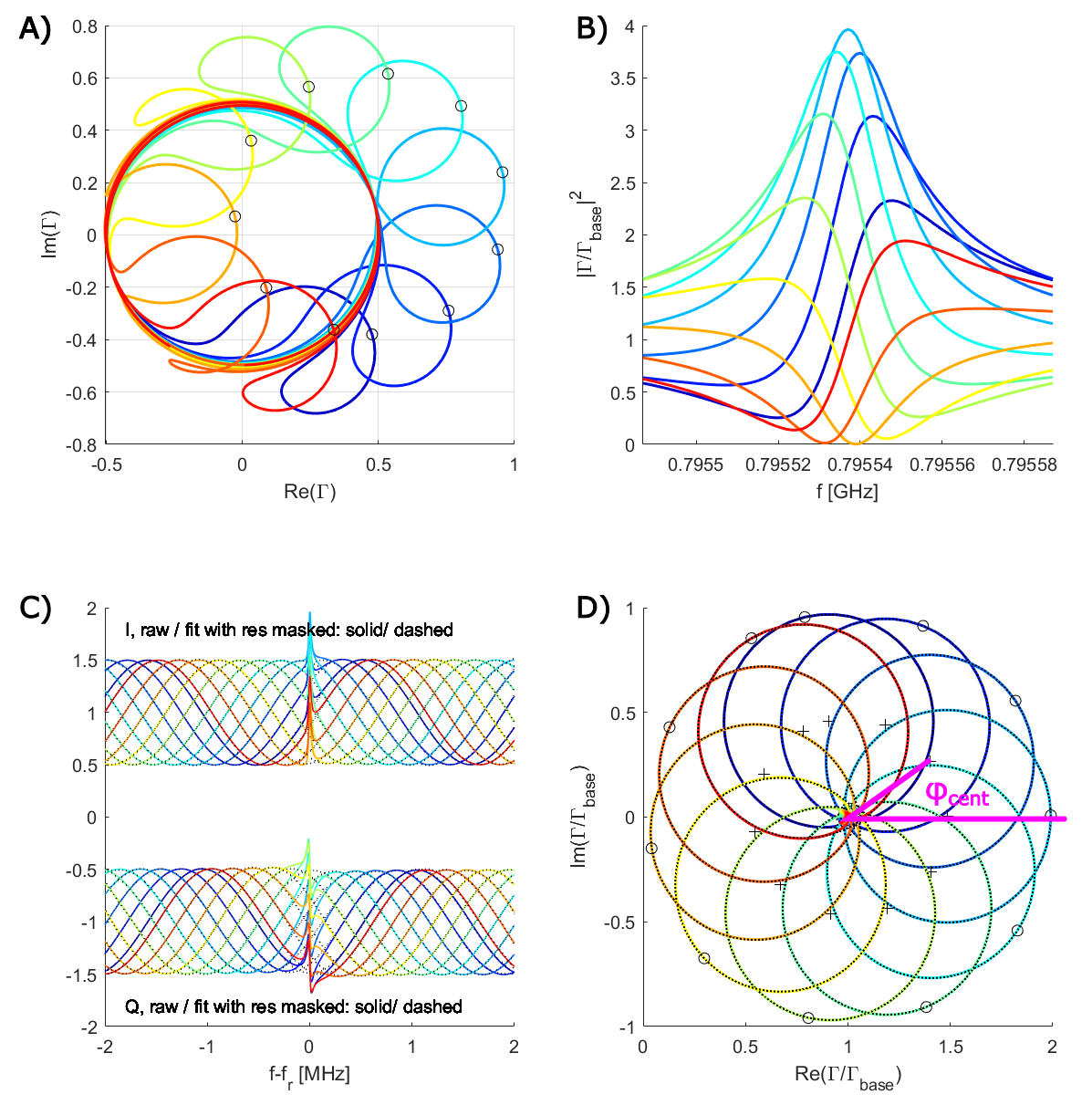}
    \caption{\textbf{A)} Simulated Raw IQ loops. Large $\tau_0$ causes several wraps around intrinsic resonator feature. Blue-red, increasing $\tau$, and just the first $\sim 2 \pi$ worth of shift is shown here. \textbf{B)} Squared magnitude w.r.t delay and frequency. The value of this on resonance, w.r.t $\tau$ also gives us interferograms.}
    \label{fig:IQeg1} \textbf{C)} I and Q w.r.t $f-f_r$, artificial vertical offsets; both I and Q modulate with current and frequency. We mask the on-resonance part and perform polynomial fits to the baseline (dashed). \textbf{D)} Calibrated I and Q. Resonator loops (dashed lines are fits and ``+'' are centers) rotate with $\tau$. Distance from (0,0) to centers, w.r.t $\tau$ gives us interferograms. Angle $\phi_{cent}(\tau)$ from off-resonance to center is the argument of the interferogram, modulo a constant. 
\end{figure}

\begin{figure*}[h!]
    \centering
    \subfloat[]{\includegraphics[width=0.45\textwidth]{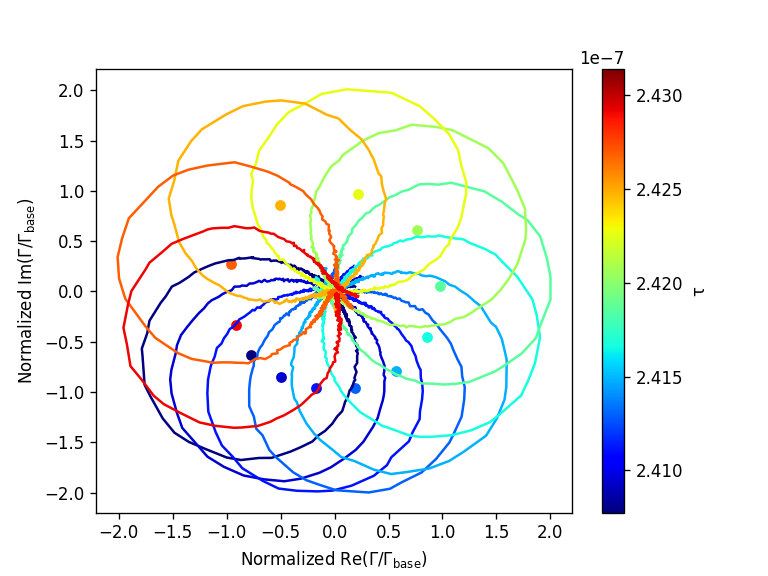} }
    \subfloat[]{\includegraphics[width=0.45\textwidth]{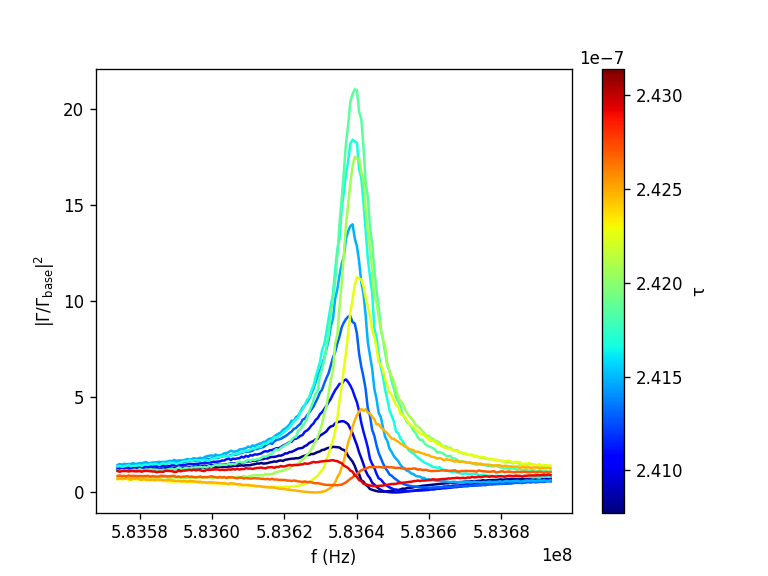} }
    \caption{ Normalized and baseline corrected IQ loop \textbf{(a)} and baseline corrected magnitude squared \textbf{(a)} as the MKID undergoes a 2$\pi$ change in phase as the bias current, and hence $\tau(I)$, is increased. Of prime importance to this work is the phase rotation of the resonance circle and \textbf{(a)} here matches Fig.~\ref {fig:IQeg1} \textbf{C)} with near perfection. The corresponding amplitude oscillations with varying $\tau$ are seen as well, though absolute calibration of impedance limit a perfect match with theory.
    }
    \label{fig:normalized_measured}
\end{figure*}

\subsection{Normalization and Baseline Calibration}

Note that, $\tau(I=0) \equiv \tau_0 = 238 \text{ns} \gg 1$ns and this has implications for phase estimation. Namely, as a function of VNA frequency, modulations exist at 0 current as $\sim \cos(4\pi f\tau_0)$. Related, on resonance, $I \sim \cos(\Omega_k D_k/v) + \cos(\Omega_k \tau_0)$ and $Q\sim \sin(\Omega_k D_k/v) + \sin(\Omega_k \tau_0)$ and $I^2 + Q^2 \neq $ constant (contrary to $\tau_0 =0$ case). We simulate the setup discussed so far and sample of $\tau$ sweeps is shown in Fig.~\ref{fig:IQeg1} A). The effect of $\tau_0 \gg $ 1 ns is seen in the winding IQ loops which when individually plotted with frequency show the modulating baselines.

We develop a calibration and normalization method to get around this, which also handles other\footnote{Some, but unlikely for all possible variations.} baseline variations, e.g., from impedance mismatches.  Polynomial fits to I and Q (dashed lines, Fig.~\ref{fig:IQeg1} B)) are used to define $\Gamma_{base}(f, \tau) = I_{fit}(f, \tau) + iQ_{fit}(f, \tau)$. Importantly, we must (i) mask the resonance and (ii) have sufficient bandwidth and fit parameters to capture the modulations. From the normalized response, there are equivalent metrics to construct interferograms:

\begin{enumerate}
 \item The angle between the center of the IQ loop ($\phi_{\text{cent}}$) rotating about the off resonance point.
 \item The distance from (0,0) (any arbitrary reference) to the IQ loops' center (obtained with circle fitting)
 \item The magnitude of the response on-resonance w.r.t delay.
\end{enumerate}

In practice, tracking $\phi_{\text{cent}}$ yielded the most useful interferograms because it was less susceptible to jitter than the on-resonance response and was more consistent across resonators than tracking distance from a reference point.

The measured signal is  
\begin{equation}
M(f, \tau) = g \left( \Gamma^{(PS)}(f,\tau) + \Gamma^{(DUT)}(f, \tau) \right)  \in \mathbf{C}
\end{equation}
where $g \approx 0.25$, like eqn.~\ref{eq:P_out}. From fitting wide scans and removing the region influenced by the resonator (fine scan), we have $M(f, \tau) \approx g \left( \Gamma^{(PS)}(f,
\tau) \right)$ i.e., the baseline. From the measurement $M$ and $\Gamma_{base}$ we can construct a normalized, baseline corrected metric $\Psi(f, \tau)$,

\begin{equation}
\label{eq:Psi_net}
    \Psi = \frac{\left(M-\Gamma_{base}\right)}{\Gamma_{base}}
\end{equation}

\section{Interferogram analysis and Results}

Each resonator is baseline calibrated and normalized, leading to agreement in the predicted vs observed resonator behavior shown in Figures \ref{fig:IQeg1} (c), (d) and \ref{fig:normalized_measured} (a), (b) respectively. 

\begin{figure}[h!]
    \centering
    \includegraphics[width=\linewidth]{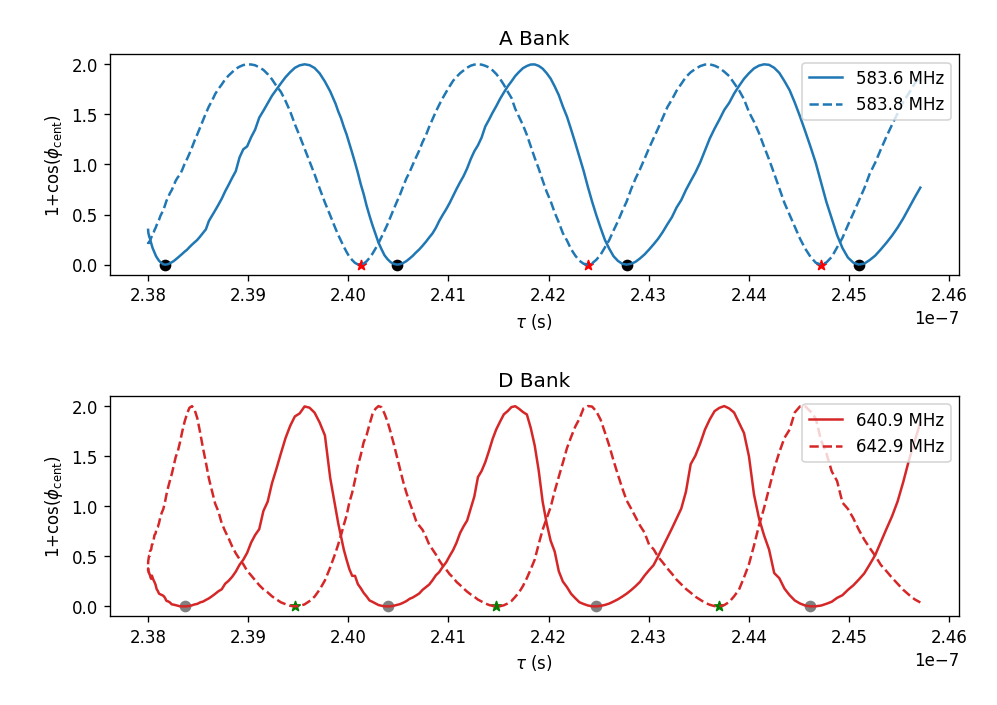} 
    \caption{ Interferograms of the four resonators that comprise the A bank (lowest frequency) and D bank, with the 583.6 MHz MKID also featured in Fig. \ref{fig:normalized_measured}. Each $\tau_n$ we observe is also demarcated.
    }
    \label{fig:interferogram}
\end{figure}

\begin{figure*}[h!]
    \centering
    \subfloat[]{\includegraphics[width=0.45\textwidth]{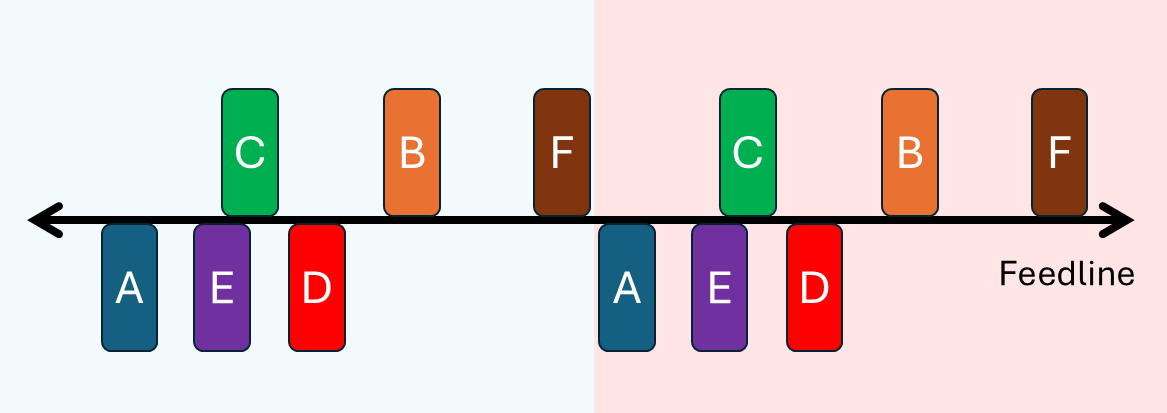} }
    \subfloat[]{\includegraphics[width=0.45\textwidth]{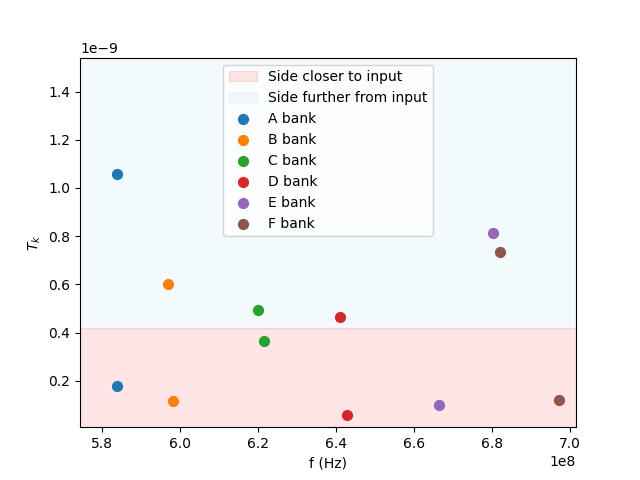} }
    \caption{ \textbf{(a)} Placement of the 6 lowest frequency banks along the array's feedline. \textbf{(b)} Calculated  electrical path lengths $T_k$ for banks A-F. The mapping does not completely reflect the layout of the array, since we would expect $T_{A} > T_{E} > T_{C} > T_{D} > T_{B} > T_{F} $ on each half of the board. However, we are still able to recover the bimodal structure of the array's design and determine which MKID is within each bank by its proximity to the input port.
    }
    \label{fig:result}
\end{figure*}

We construct an interferogram from $\phi_{\text{cent}}$ of the corrected resonator and fit it with $\text{Fit}(\tau) = p_o + p_a \cos(p_b\tau + p_c)$ to extract the period. At the $n^{th}$ minimums of the interferogram, the argument of the cosine can be extracted as $p_b\tau_n + p_c = (2n+1)\pi$. Using the fitted periodicity for resonator $k$, we can estimate the first minimum, $\tau_{k,0}$, to solve for the feedline length to that resonator.

\begin{equation}
    T_k = \tau_{k,0} - \frac{1}{\Omega_k} (\phi_0 + \pi)
\end{equation}

This presents three possible cases. If $\phi_0$=0, we can solve each equation directly for $T_k$. If $\phi_0$ is an unknown global constant, then we can solve the system of equations

\begin{equation}
    \{T_k = \frac{f_1}{f_k} (T_1 -\tau_{1,0}) +\tau_{k,0}\}_{k=2,...,K},
\end{equation}

where $T_1$ is estimated using the designed frequency schedule of the array and the phase velocity in the feedline.

In the third case, $\phi_0 = \phi_{k,0}$ is a distinct zero-current phase for each resonator. To handle this requires 2 measurements, where the setup is kept the same except the DUT is flipped so that its input and terminated ports swap. Thus, if the total electrical length of the feedline is $T_{\text{tot}}$ and the two measured delays are denoted $\tau_{k,0}, \tau_{k,0}'$, then 

\begin{equation}
\begin{cases}
T_k - \tau_{k,0} = \frac{\phi_{k,0} + \pi}{\Omega_k} \\
(T_{\text{tot}}-T_k)  - \tau_{k,0}' = \frac{\phi_{k,0} + \pi}{\Omega_k}  
\end{cases}
\end{equation}

\begin{equation}
T_k = \frac{1}{2} \left( T_{\text{tot}}+\tau_{k,0}-\tau_{k,0}' \right)
\end{equation}

Currently, the DUT has only been fully analyzed in a single orientation. The array layout is divided into 22 banks of MKIDs, each bank consisting of a pair of resonators nearby in frequency. Each half of the array has a pixel from each bank. This bisection is observed in figure \ref{fig:result}, but the bank ordering within each half does not match the design. These results suggest that $\phi_0$ is not a global constant on this device.

\section{Future work}

The flipped orientation analysis is on-going. Initial analysis yields similar conclusions to the single orientation mode, but there is ample room for improvement in the baseline correction step. The current focus of study is tracking the baseline as a function of bias current to be able to smoothly model it in both frequency and bias current, as opposed to creating a new instance of it for each bias value. Another potential improvement in the DUT modeling would be to measure the variation of wave speed $v = 1/\sqrt{\mathcal{L}\mathcal{C}}$ across the length of the feedline by probing kinetic inductance $\mathcal{L}_k$ at the location of each resonator by examining the feedline-resonator impedance mismatch.

This manner of superconducting interferometry has been developed at microwave frequencies for Superconducting On-chip Fourier Transform Spectrometers (SOFTS)~\cite{BT20, BT23}. In future work, we may leverage SOFTS which has customizable broadband QHCs and integrated phase tunable transmission lines to map MKID arrays. This will not only alleviate issues relating to matching impedances and bandwidths between commercial elements and the MKID arrays, but also allow for a much more compact setup.

\section*{Acknowledgment}

The research was carried out at the Jet Propulsion Laboratory, California Institute of Technology, under a contract with the National Aeronautics and Space Administration (80NM0018D0004). This work was supported by NASA Space Technology Graduate Research Opportunities (NSTGRO) program under grant number 80NSSC24K1395 to Chris Albert and by the NASA Strategic Astrophysics Technology (SAT) program under grant number 141108.04.02.01.70 to C.M. Bradford et al. 


\end{document}